\documentclass[conference]{IEEEtran}
\IEEEoverridecommandlockouts
\usepackage{cite}
\usepackage{amsmath,amssymb,amsfonts}
\usepackage{algorithmic}
\usepackage{graphicx}
\usepackage{textcomp}
\usepackage{xcolor}
\def\BibTeX{{\rm B\kern-.05em{\sc i\kern-.025em b}\kern-.08em
    T\kern-.1667em\lower.7ex\hbox{E}\kern-.125emX}}

\usepackage[hidelinks]{hyperref}
\usepackage{cite}

\usepackage{subfigure}

\renewcommand{\Vec}{\mathbf}
\usepackage{siunitx}

\usepackage{empheq}

\begin{document}

\title{CASCADED NOISE REDUCTION AND ACOUSTIC ECHO CANCELLATION BASED ON AN EXTENDED NOISE REDUCTION
\thanks{This research was carried out at the ESAT Laboratory of KU Leuven, in the frame of Research Council KU Leuven C14-21-0075 "A holistic approach to the design of integrated and distributed digital signal processing algorithms for audio and speech communication devices", and Aspirant Grant 11PDH24N (for A. Roebben) from the Research Foundation - Flanders (FWO).}
}

\author{\IEEEauthorblockN{Arnout Roebben, Toon van Waterschoot, and Marc Moonen}\\
\IEEEauthorblockA{\textit{Department of Electrical Engineering (ESAT)}\\ \textit{STADIUS Center for Dynamical Systems, Signal Processing and Data Analytics}} \\
\textit{KU Leuven}\\
Leuven, Belgium \\
\{arnout.roebben, toon.vanwaterschoot, marc.moonen\}@kuleuven.be}

\maketitle
 
\begin{abstract}
 In many speech recording applications, the recorded desired speech is corrupted by both noise and acoustic echo, such that combined noise reduction (NR) and acoustic echo cancellation (AEC) is called for. A common cascaded design corresponds to NR filters preceding AEC filters. These NR filters aim at reducing the near-end room noise (and possibly partially the echo) and operate on the microphones only, consequently requiring the AEC filters to model both the echo paths and the NR filters. In this paper, however, we propose a design with extended NR (NR\textsubscript{ext}) filters preceding AEC filters under the assumption of the echo paths being additive maps, thus preserving the addition operation. Here, the NR\textsubscript{ext} filters aim at reducing both the near-end room noise and the far-end room noise component in the echo, and operate on both the microphones and loudspeakers. We show that the succeeding AEC filters remarkably become independent of the NR\textsubscript{ext} filters, such that the AEC filters are only required to model the echo paths, improving the AEC performance. Further, the degrees of freedom in the NR\textsubscript{ext} filters scale with the number of loudspeakers, which is not the case for the NR filters, resulting in an improved NR performance.
\end{abstract}


\section{Introduction} \label{section:introduction}
\IEEEPARstart{I}{n} many speech recording applications, such as hands-free telephony in cars or hearing instruments, the recorded desired speech is corrupted by both noise and acoustic echo. This noise originates from within the room, the so-called near-end room, whereas the echo originates from loudspeakers playing signals recorded in another room, the so-called far-end room. To remove the noise and echo, combined noise reduction (NR) and acoustic echo cancellation (AEC) is called for. 

Acoustic echo cancellation (AEC) aims at removing the echo while preserving the desired speech. AEC algorithms traditionally exploit the loudspeaker signals to compute an estimate of this echo, which can subsequently be subtracted from the microphone signals \cite{hanslerTopicsAcousticEcho2006}. On the other hand, noise reduction (NR) aims at removing the near-end room noise while preserving the desired speech \cite{serizelLowrankApproximationBased2014}.

Algorithms for combined AEC and NR aim at jointly removing the echo and near-end room noise while preserving the desired speech \cite{gustafssonCombinedAcousticEcho1998a,docloCombinedAcousticEcho2000,hanslerTopicsAcousticEcho2006}. Generally, a cascaded approach is adhered to, where either the AEC precedes the NR (AEC-NR) \cite{gustafssonCombinedAcousticEcho1998a,cohenJointBeamformingEcho2018,luisvaleroLowComplexityMultiMicrophoneAcoustic2019a}, or the NR precedes the AEC (NR-AEC) \cite{martinAcousticEchoCancellation1997,docloCombinedAcousticEcho2000,schrammenChangePredictionLow2019}. NR-AEC is beneficial computational complexity-wise, only requiring one AEC filter for each loudspeaker rather than for each microphone-loudspeaker pair. Further, the AEC filters in NR-AEC operate under reduced noise influence. However, as the microphone signals in NR-AEC are affected by the NR filters, the AEC filters do not solely model the echo paths between the loudspeakers and microphones, but rather the combination of echo paths and NR filters \cite{reuvenJointAcousticEcho2004,docloCombinedAcousticEcho2000,luisvaleroLowComplexityMultiMicrophoneAcoustic2019a}.

In this paper, to combat this drawback, we propose a design where an extended NR (NR\textsubscript{ext}) precedes the AEC (NR\textsubscript{ext}-AEC) under the assumption of the echo paths being additive maps, thus preserving the addition operation. Whereas the NR filters aim at reducing the near-end room noise (and possibly partially the echo) and operate on the microphones only, the NR\textsubscript{ext} filters aim at reducing both the near-end room noise and the far-end room noise component in the echo, and operate on both the microphones and loudspeakers. We show that the AEC filters remarkably become independent of the NR\textsubscript{ext} filters, such that the AEC filters are thus only required to model the echo paths, thereby improving the AEC performance. Additionally, the NR\textsubscript{ext} filters attain a better NR performance as the degrees of freedom in the NR\textsubscript{ext} filters scale with the number of loudspeakers, which is not the case for the NR filters. This improved performance is demonstrated by means of simulations, of which the code is available at \cite{roebbenGithubRepositoryCascaded2024}. 

A design similar to NR\textsubscript{ext}-AEC has previously been proposed for the related, but distinct, problem of combined NR and acoustic feedback cancellation (AFC) \cite{ruizCascadeAlgorithmsCombined2023a}. However, to the best of our knowledge, it has not been proposed for combined NR and AEC, which in particular leads to the additional observation proven in this paper that the AEC filters become independent of the NR\textsubscript{ext} filters.

\section{Signal model} \label{section:signal_model}
The signal model is presented in the $z$-domain to accommodate the duality between frequency- and time-domain. The $z$- to frequency-domain conversion is realised by replacing index $z$ with frequency-bin index $f$ (and possibly frame index $k$). The $z$- to time-domain conversion is realised by replacing the $z$-domain variables with time-lagged vectors. 

Considering an $L$-loudspeaker/$M$-microphone setup, the microphone signals $m_m(z)$, $m\in\{1,\dots, M\}$, can be stacked into the microphone signal vector $\Vec{m}(z)\in\mathbb{C}^{M\times 1}$ as
\begin{equation} \label{eq:signal_model_def_m_z_domain}
	\Vec{m}(z) = \begin{bmatrix}
		m_1(z)&
		m_2(z)&
		\dots&
		m_M(z)
	\end{bmatrix}^\top_{\textstyle \raisebox{2pt}{,}}
\end{equation}
and decomposed into the desired speech signal vector $\Vec{s}(z)\in\mathbb{C}^{M\times 1}$, the near-end room noise signal vector $\Vec{n}(z)\in\mathbb{C}^{M\times 1}$, and the echo signal vector $\Vec{e}(z)\in\mathbb{C}^{M\times 1}$ as 
\begin{equation} \label{eq:signal_model_AEC_s+u}
	\Vec{m}(z) =\Vec{s}(z) + \Vec{n}(z) + \Vec{e}(z)_{\textstyle \raisebox{2pt}{.}}
\end{equation}

Similarly, the loudspeaker signals, $l_l(z)$, $l\in\{1,\dots, L\}$, can be stacked into the loudspeaker signal vector $\Vec{l}(z)\in\mathbb{C}^{L\times 1}$ as
\begin{equation} \label{eq:signal_model_def_l_z_domain}
	\Vec{l}(z) = \begin{bmatrix}
		l_1(z)&
		l_2(z)&
		\dots&
		l_L(z)
	\end{bmatrix}^\top_{\textstyle \raisebox{2pt}{,}}
\end{equation}
and split in a far-end room speech component $\Vec{l}^s(z)$ and far-end room noise component $\Vec{l}^n(z)$, i.e., $\Vec{l}(z) = \Vec{l}^s(z) + \Vec{l}^n(z)$, thus modelling far-end room speech and noise respectively.

We assume the echo paths to be additive maps $F(.):\mathbb{C}^{L\times 1}\to\mathbb{C}^{M\times 1}$, where $\Vec{e}(z)=F\left(\Vec{l}(z)\right)$ can thus be decomposed into a speech component in the echo $\Vec{e}^s(z)=F\left(\Vec{l}^s(z)\right)$ and a noise component in the echo $\Vec{e}^n(z)=F\left(\Vec{l}^n(z)\right)$, i.e.,
\begin{equation}
		\Vec{e}(z) =F\left(\Vec{l}^s(z)\right)+F\left(\Vec{l}^n(z)\right)= \Vec{e}^s(z) + \Vec{e}^n(z)_{\textstyle \raisebox{2pt}{.}}
\end{equation} 
The addition operation is thus preserved by the echo paths. An example of an echo path satisfying this additivity assumption is a linear echo path, possibly followed by sub-Nyquist sampling. 

In addition, leveraging both the available microphone and loudspeaker signal vectors, an extended microphone signal vector $\tilde{\Vec{m}}(z)\in\mathbb{C}^{\left(M+L\right)\times 1}$ can be defined as
\begin{subequations} \label{eq:signal_model_def_x_tilde}
	\begin{align}
		\tilde{\Vec{m}}(z) &= \begin{bmatrix} \Vec{m}(z)^\top & \Vec{l}(z)^\top \end{bmatrix}^\top\\
		&= \tilde{\Vec{s}}(z)+\tilde{\Vec{n}}(z) + \tilde{\Vec{e}}^s(z)+\tilde{\Vec{e}}^n(z)\\
		&= \begin{bmatrix} \Vec{s}(z) \\ \Vec{0}_{L\times1} \end{bmatrix}+\begin{bmatrix} \Vec{n}(z) \\ \Vec{0}_{L\times1} \end{bmatrix}+\begin{bmatrix} \Vec{e}^s(z) \\ \Vec{l}^s(z) \end{bmatrix}+\begin{bmatrix} \Vec{e}^n(z) \\ \Vec{l}^n(z) \end{bmatrix}_{\textstyle \raisebox{2pt}{.}}
	\end{align}
\end{subequations}	

We also make the following assumptions:
\begin{itemize}
	\item $\Vec{s}(z)$, $\Vec{n}(z)$, $\Vec{e}^s(z)$ and $\Vec{e}^n(z)$ are uncorrelated, $\Vec{l}^s(z)$ and $\Vec{l}^n(z)$ are uncorrelated, and $\Vec{l}^s(z)$ and $\Vec{l}^n(z)$ are uncorrelated with $\Vec{s}(z)$ and $\Vec{n}(z)$.
	\item The microphone correlation matrix $R_{mm}(z)=\mathbb{E}\left\{\Vec{m}(z)\Vec{m}(z)^H\right\}=R_{ss}(z)+R_{nn}(z)+R_{ee}(z)\in\mathbb{C}^{M\times M}$ is of full rank, with $R_{ss}(z)=\mathbb{E}\left\{\Vec{s}(z)\Vec{s}(z)^H\right\}$, $R_{nn}(z)=\mathbb{E}\left\{\Vec{n}(z)\Vec{n}(z)^H\right\}$ and $R_{ee}(z)=\mathbb{E}\left\{\Vec{e}(z)\Vec{e}(z)^H\right\}$. Here, $R_{ee}(z)$ can be decomposed into $R_{e^se^s}(z)=\mathbb{E}\left\{\Vec{e}^s(z)\Vec{e}^s(z)^H\right\}$ and $R_{e^ne^n}(z)=\mathbb{E}\left\{\Vec{e}^n(z)\Vec{e}^n(z)^H\right\}$. The loudspeaker correlation matrix $R_{ll}(z)=\mathbb{E}\left\{\Vec{l}(z)\Vec{l}(z)^H\right\}=R_{l^sl^s}(z)+R_{l^nl^n}(z)\in\mathbb{C}^{L\times L}$ is of full rank, with $R_{l^sl^s}(z)=\mathbb{E}\left\{\Vec{l}^s(z)\Vec{l}^s(z)^H\right\}$ and $R_{l^nl^n}(z)=\mathbb{E}\left\{\Vec{l}^n(z)\Vec{l}^n(z)^H\right\}$. Similarly, $R_{l^sl^s}(z)$ is of full rank, e.g., due to multiple far-end room speech sources or due to non-linearities that are additive maps in the transmission from far-end to near-end room. The general case where the correlation matrices can be rank-deficient will not be considered here for the sake of conciseness but will be considered elsewhere.
	\item The mean squared error (MSE) optimal estimates of the echo paths between the loudspeakers and microphones are equal for the echo and the far-end room speech component in the echo, i.e., $R_{ll}(z)^{-1} R_{le}(z)=R_{l^sl^s}(z)^{-1} R_{l^se^s}(z)$. Here, $R_{le}(z)=\mathbb{E}\left\{\Vec{l}(z)\Vec{e}(z)^{H}\right\} \in\mathbb{C}^{L\times M}$, and $R_{l^se^s}(z)=\mathbb{E}\left\{\Vec{l}^s(z)\Vec{e}^{s}(z)^H\right\}$.
\end{itemize}
Similarly, the extended microphone correlation matrix can be defined as $R_{\tilde{m}\tilde{m}}=\mathbb{E}\{\tilde{\Vec{m}}(z)\tilde{\Vec{m}}(z)^H\}=R_{\tilde{s}\tilde{s}}(z)+R_{\tilde{n}\tilde{n}}(z)+R_{\tilde{e}\tilde{e}}(z)\in\mathbb{C}^{(M+L)\times (M+L)}$ with $R_{\tilde{s}\tilde{s}}=\mathbb{E}\{\tilde{\Vec{s}}(z)\tilde{\Vec{s}}(z)^H\}$, $R_{\tilde{n}\tilde{n}}=\mathbb{E}\{\tilde{\Vec{n}}(z)\tilde{\Vec{n}}(z)^H\}$, and $R_{\tilde{e}\tilde{e}}=\mathbb{E}\{\tilde{\Vec{e}}(z)\tilde{\Vec{e}}(z)^H\}$. Here, $R_{\tilde{e}\tilde{e}}(z)$ can be decomposed into $R_{\tilde{e}^s\tilde{e}^s}(z)=\mathbb{E}\{\tilde{\Vec{e}}^s(z)\tilde{\Vec{e}}^s(z)^H\}$ and $R_{\tilde{e}^n\tilde{e}^n}(z)=\mathbb{E}\{\tilde{\Vec{e}}^n(z)\tilde{\Vec{e}}^n(z)^H\}$.

While $\Vec{n}(z)$ and $\Vec{e}^n(z)$ can be assumed to be stationary always-on signal vectors, $\Vec{s}(z)$ and $\Vec{e}^s(z)$ are non-stationary on-off signal vectors. A voice activity detector (VAD) is assumed available to distinguish between these on-off periods. To this end, $\text{VAD}_{s}(z)$ differentiates between on-off periods in $\Vec{s}(z)$, and $\text{VAD}_{e^s}(z)$ in $\Vec{e}^s(z)$. The following regimes can be defined:
\begin{itemize}[\IEEElabelindent=0.1em]
	\item $\text{VAD}_{s}(z)=1$, $\text{VAD}_{e^s}(z)=1$: $\Vec{m}(z)=\Vec{s}(z)+\Vec{n}(z)+\Vec{e}(z)$;
	\item $\text{VAD}_{s}(z)=1$, $\text{VAD}_{e^s}(z)=0$: $\Vec{m}(z)=\Vec{s}(z)+\Vec{n}(z)+\Vec{e}^n(z)$;
	\item $\text{VAD}_{s}(z)=0$, $\text{VAD}_{e^s}(z)=1$: $\Vec{m}(z)=\Vec{n}(z)+\Vec{e}(z)$;
	\item $\text{VAD}_{s}(z)=0$, $\text{VAD}_{e^s}(z)=0$: $\Vec{m}(z)=\Vec{n}(z)+\Vec{e}^n(z)$.
\end{itemize}

\section{Filters} \label{section:algorithms}
\begin{figure}
	\centering
	\subfigure[NR-AEC]{
		\includegraphics[width=0.46\linewidth]{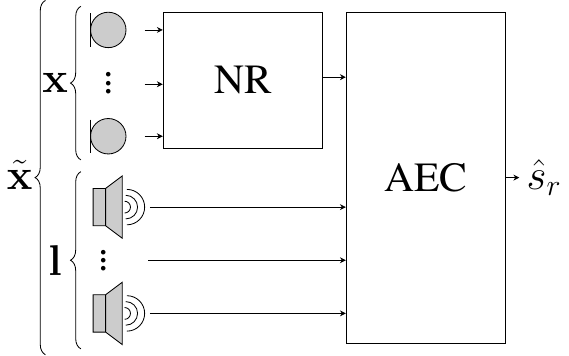}
		\label{fig:NR-AEC_overview}
	}
	\centering
	\subfigure[NR\textsubscript{ext}-AEC]{
		\includegraphics[width=0.46\linewidth]{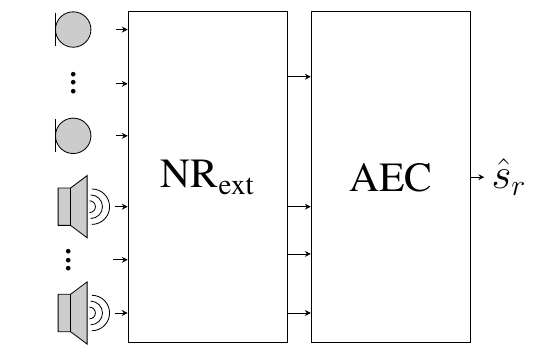}
		\label{fig:NRext-AEC_overview}
	}
	\caption{(a) The NR filters aim at reducing the near-end room noise (and possibly partially the echo), and the AEC filters aim at reducing the echo. (b) The NR\textsubscript{ext} filters aim at reducing both the near-end room noise and far-end room noise component in the echo, and the AEC filters aim at reducing the far-end room speech (and residual noise) component in the echo.}
	\label{fig:schematics_overview}
\end{figure}

Combined NR and AEC aims at designing a filter $\tilde{\Vec{w}}_r(z)\in\mathbb{C}^{\left(M+L\right)\times 1}$ to estimate the desired speech in reference microphone $r\in\{1,\cdots,M\}$. As illustrated in Fig. \ref{fig:NR-AEC_overview}, commonly, $\tilde{\Vec{w}}_r(z)$ is designed by cascading NR filters to reduce the near-end room noise (and possibly partially the echo) in the microphones with AEC filters to reduce the echo. This design, referred to as NR-AEC, is discussed in Section \ref{section:algorithms-NRAEC}.

To combat dependence of the AEC filters on the NR filters in NR-AEC, we propose to precede the AEC filters with NR\textsubscript{ext} filters as illustrated in Fig. \ref{fig:NRext-AEC_overview} and referred to as NR\textsubscript{ext}-AEC. Contrary to NR-AEC, the AEC filters remarkably become independent of the NR\textsubscript{ext} filters as discussed in Section \ref{section:algorithms-NRextAEC}.

\subsection{NR first, AEC second (NR-AEC)} \label{section:algorithms-NRAEC}
In NR-AEC, $\tilde{\Vec{w}}_r(z)$ is designed to reconstruct the desired speech in the $r$th microphone as $\hat{s}_r(z)=\tilde{\Vec{w}}_r(z)^H\tilde{\Vec{m}}(z)$ with: 
\begin{subequations} \label{eq:NRAEC}
		\begin{empheq}[box=\fbox]{align}
			&\tilde{\Vec{w}}_r(z) = \underbrace{\begin{bmatrix}
					W_{NR}(z) & 0_{M\times L}\\
					0_{L\times M} & \mathbb{I}_{L\times L}
			\end{bmatrix}}_{NR}\underbrace{\begin{bmatrix}
					\mathbb{I}_{M\times M}\\ -W_{AEC}^\star(z)
			\end{bmatrix}}_{AEC}\tilde{\Vec{t}}_{r_{\textstyle \raisebox{2pt}{,}}} \text{ with}\\
				& W_{NR}(z) = R_{mm}(z)^{-1}R_{ss}(z)\\
				& W_{AEC}^\star(z) = R_{ll}(z)^{-1}R_{le^{\star}}(z)_{\textstyle \raisebox{2pt}{.}}
		\end{empheq}
\end{subequations}
Here, $\Vec{e}^{\star}(z)$ is defined as $W_{NR}(z)^H\Vec{e}(z)$, i.e., as the echo signal vector after applying $W_{NR}(z)$, and $R_{le^{\star}}(z)=\mathbb{E}\{\Vec{l}(z)\Vec{e}^{\star}(z)^H\}$ corresponds to the loudspeaker-echo correlation matrix after applying $W_{NR}(z)$. $\tilde{\Vec{t}}_r\in\mathbb{C}^{(M+L)\times 1}$ is a unit vector with $1$ at position $r$ and $0$ elsewhere.

$W_{NR}(z)$ is a multichannel Wiener filter (MWF) that operates on the microphones, aimed at suppressing the near-end room noise (and echo) while preserving the desired speech \cite[Section 2.4]{sprietAdaptiveFilteringTechniques2004}. $W^{\star}_{AEC}(z)$ corresponds to an MSE optimal estimate of the echo paths after applying $W_{NR}(z)$, by means of which the echo is suppressed \cite[Section 2.1]{romboutsAdaptiveFilteringAlgorithms2003}. As the AEC filters are applied after the NR filters, the AEC filters here do not solely model the echo paths, but rather the combination of the echo paths and the NR filters, i.e., $R_{ll}(z)^{-1}R_{le^{\star}}(z) \neq R_{ll}(z)^{-1}R_{le}(z)$.

In \cite{docloMultimicrophoneNoiseReduction2003}, (\ref{eq:NRAEC}) has been considered in an $L$-loudspeaker/$1$-microphone setup by assuming linear echo paths and a sufficiently long $\tilde{\Vec{w}}_r(z)$ to model the echo paths. Under these conditions, (\ref{eq:NRAEC}) was found optimal in MSE sense, except that $R_{mm}(z)$ should be replaced with $R_{ss}(z)+R_{nn}(z)$. $W_{NR}(z)$ is then a true NR filter only aimed at reducing the near-end room noise \cite{docloMultimicrophoneNoiseReduction2003}. Nevertheless, $R_{mm}(z)$ is used both in \cite{docloMultimicrophoneNoiseReduction2003} and in this paper, as $R_{ss}(z)+R_{nn}(z)$ cannot be estimated due to $\Vec{e}(z)$ being always-on. A similar reasoning can be applied to an $L$-loudspeaker/$M$-microphone scenario, with undermodelling of the echo paths, but is omitted here for conciseness.

\subsection{NR\textsubscript{ext} first, AEC second (NR\textsubscript{ext}-AEC)} \label{section:algorithms-NRextAEC}
In NR\textsubscript{ext}-AEC, $\tilde{\Vec{w}}_r(z)$ is designed to reconstruct the desired speech in the $r$th microphone as $\hat{s}_r(z)=\tilde{\Vec{w}}_r(z)^H\tilde{\Vec{m}}(z)$ with: 
\begin{subequations} \label{eq:NRextAEC}
\begin{empheq}[box=\fbox]{align} 
			&\tilde{\Vec{w}}_r(z) = \underbrace{W_{NR_{ext}}(z)}_{NR_{ext}}\underbrace{\begin{bmatrix}
					\mathbb{I}_{M\times M} \\-W_{AEC}^{\star\star} (z)
			\end{bmatrix}}_{AEC}\tilde{\Vec{t}}_{r_{\textstyle \raisebox{2pt}{,}}} \text{ with}\\
		& W_{NR_{ext}}(z) = R_{\tilde{m}\tilde{m}}(z)^{-1}\left(R_{\tilde{s}\tilde{s}}(z)+R_{\tilde{e}^s\tilde{e}^s}(z)\right)\\
		& W_{AEC}^{\star\star}(z) = R_{l^{\star\star}l^{\star\star}}(z)^{-1} R_{l^{\star\star}m^{\star\star}}(z)_{\textstyle \raisebox{2pt}{.}}
	\end{empheq}
\end{subequations}
Here, $\Vec{m}^{\star\star}(z)$ and $\Vec{l}^{\star\star}(z)$ are defined as $\begin{bmatrix}
	\Vec{m}^{\star\star}(z)^\top & \Vec{l}^{\star\star}(z)^\top
\end{bmatrix}^\top=W_{NR_{ext}}(z)^H\tilde{\Vec{m}}(z)$, i.e., as the microphone and loudspeaker signal vectors after applying $W_{NR_{ext}}(z)$, and $R_{l^{\star\star}l^{\star\star}}(z)=\mathbb{E}\{\Vec{l}^{\star\star}(z)\Vec{l}^{\star\star}(z)^H\}$ and $R_{l^{\star\star}m^{\star\star}}(z)=\mathbb{E}\{\Vec{l}^{\star\star}(z)\Vec{m}^{\star\star}(z)^H\}$.

$W_{NR_{ext}}(z)$ is an MWF that operates on the extended signal model, aimed at suppressing the near-end room noise and far-end room noise component in the echo while preserving the desired speech and far-end room speech component in the echo. $W_{AEC}^{\star\star}(z)$ corresponds to an MSE optimal estimate of the echo paths between the loudspeakers and microphones, by means of which the far-end room speech component in the echo and the residual far-end room noise component in the echo are suppressed. Although the AEC filters are computed after the NR\textsubscript{ext} filters, contrary to NR-AEC, the AEC filters are not affected by the NR\textsubscript{ext} filters. Indeed, rewriting $W_{NR_{ext}}(z)$ using the $2\times 2$ block inverse formula \cite[(2.3)]{luInversesBlockMatrices2002}, $R_{l^{\star\star}l^{\star\star}}(z)$ and $R_{l^{\star\star}m^{\star\star}}(z)$ are given as:
\begin{subequations} \label{eq:NRext_AECfilters}
	\begin{align}
		&R_{l^{\star\star}l^{\star\star}}(z) = R_{l^sl^s}(z)R_{ll}(z)^{-1}R_{l^sl^s}(z)\\
		&R_{l^{\star\star}m^{\star\star}}(z) = R_{l^sl^s}(z)R_{ll}(z)^{-1}R_{l^se^s}(z)_{\textstyle \raisebox{2pt}{.}}
	\end{align}
\end{subequations}
Using (\ref{eq:NRext_AECfilters}), $W^{\star\star}_{AEC}(z)$ corresponds to:
\begin{equation}
	W^{\star\star}_{AEC}(z) = R_{l^sl^s}(z)^{-1}R_{{l^se^s}}(z)_{\textstyle \raisebox{2pt}{,}}
\end{equation}
which is indeed the MSE optimal estimate of the echo paths, and so remarkably independent of the NR\textsubscript{ext} filters. The optimality of (\ref{eq:NRextAEC}) will be discussed elsewhere.

\section{Experimental procedures} \label{section:experimental_procedures}
\subsection{Acoustic scenarios} \label{section:experimental_procedures-acoustic_scenarios}
In a $\SI{5}{\meter}$ $\times$ $\SI{5}{\meter}$ $\times$ $\SI{3}{\meter}$ room with a reflection coefficient of $0.15$, source-to-mic and loudspeaker-to-mic impulse responses of length $128$ samples at a sampling rate of $\SI{16}{\kilo\hertz}$ are generated using the randomised image method (RIM) with randomised distances of $\SI{0.13}{\meter}$ \cite{desenaModelingRectangularGeometries2015}. One desired speech source, one near-end room noise source and $L=2$ loudspeakers are placed at congruent angles in a circle with a $\SI{0.2}{\meter}$ radius around the mean microphone position. $M=2$ microphones are positioned at $\begin{bmatrix} 2 & 1.9 & 1\end{bmatrix}$$\SI{}{\meter}$ and $\begin{bmatrix} 2 & 1.8 & 1 \end{bmatrix}$$\SI{}{\meter}$. Five scenarios with varying desired speech source, near-end room noise source and loudspeaker positions are examined.

Sentences of the hearing in noise test (HINT) database, concatenated with $\SI{5}{\second}$ of silence, are used as a desired speech source and far-end room speech component in the loudspeakers \cite{nilssonDevelopmentHearingNoise1994}. Babble noise is used as the near-end room noise source to model competing talkers \cite{auditecAuditoryTestsRevised1997}, and white noise as the far-end room component in the loudspeakers to model, e.g., sensor and far-end room noise. All signals are $\SI{10}{\second}$ long. 

The power ratio between the echo signals is set to $\SI{0}{\decibel}$. The power ratio between $\Vec{l}^s(z)$ and $\Vec{l}^n(z)$ equals $\SI{0}{\decibel}$. The input signal-to-noise ratio (SNR\textsuperscript{in}) and signal-to-echo (SER\textsuperscript{in}) ratio in microphone $r$ are varied between $\SI{-15}{\decibel}$ and $\SI{15}{\decibel}$. 

\subsection{Algorithm settings} \label{section:experimental_procedures-algorithmic_settings}
The filters $W_{NR}(z)$ and $W_{NR_{ext}}(z)$ are implemented in the short-time Fourier transform- (STFT) domain as frequency-domain filters attain larger SNR improvements than time-domain filters \cite[Chapter 2]{sprietAdaptiveFilteringTechniques2004}. To this end, a squared root Hann window of size $512$ samples with a window shift of $256$ samples is used. The correlation matrices are estimated by averaging across frames. As one desired speech source is considered, the rank of the desired speech correlation matrix at frequency-bin $f$, $R_{ss}(f)$, is enforced to equal one by using a generalised eigenvalue decomposition (GEVD) \cite{serizelLowrankApproximationBased2014}. Similarly, as one desired speech source and two independent loudspeaker sources are considered, the rank of $R_{\tilde{s}\tilde{s}}(f)+R_{\tilde{e}^s\tilde{e}^s}(f)$ is enforced to equal three by using a GEVD. The NR and NR\textsubscript{ext} filters are nevertheless applied in the time-domain by converting the STFT-domain filters to an equivalent distortion function of $2N-1$ samples \cite{didierSamplingRateOffset2023}, as the AEC filters need to model the NR filters and frequency-domain filters cannot be modelled exactly by the AEC filters \cite{avargelSystemIdentificationShortTime2007a}.

The AEC filters $W_{AEC}^{\star}(z)$ and $W_{AEC}^{\star\star}(z)$ are implemented in the time-domain as STFT-domain filters cannot model the echo path impulse responses exactly \cite{avargelSystemIdentificationShortTime2007a}. To this end, a normalised least mean squares (NLMS) implementation is used with step size $0.1$ and regularisation $10^{-6}$ \cite[Section 2.2]{romboutsAdaptiveFilteringAlgorithms2003}. To reduce excess error in the NLMS updates, the AEC filters are updated only when the desired speech is inactive. 

For NR-AEC, the number of coefficients in the AEC filter between the $r$th microphone and each loudspeaker $L_{\hat{F}}$ is varied between the length of the echo path impulse responses ($128$) and the length of the convolution of the NR filters and the echo path impulse responses ($(2\cdot 512-1) + 128 -1=1150$). For NR\textsubscript{ext}-AEC, the coefficients exceeding coefficient index $128$ are explicitly set to $0$ as the AEC filters only aim at modelling the echo path impulse responses.

First, to study the converged filters, the correlation matrices in the NR and NR\textsubscript{ext} filters are calculated across the entire data, and the final NLMS updated AEC filters are used for the entire data. Second, the filters are adapted through time. For the NR and NR\textsubscript{ext} filters, to this end, exponential smoothing with a weight for the previous estimate of $0.995$ is used. Ideal VADs are assumed to isolate the influence of VAD errors, but practical implementations can be found, e.g., in \cite{zhuRobustLightweightVoice2023b}. 

\subsection{Performance measures} \label{section:experimental_procedures-performance_measures}
Performance is evaluated using the intelligibility-weighted SNR improvement $\Delta \text{SNR\textsuperscript{I}}$, the intelligibility-weighted SER improvement $\Delta \text{SER\textsuperscript{I}}$ and the intelligibility-weighted speech distortion (SD) $\text{SD\textsuperscript{I}}$ \cite{greenbergIntelligibilityWeightedMeasures1993,sprietAdaptiveFilteringTechniques2004}. 

\section{Results and discussion} \label{section:results_and_discussion}
\begin{figure}	
	\centering
	\includegraphics[width=0.98\linewidth]{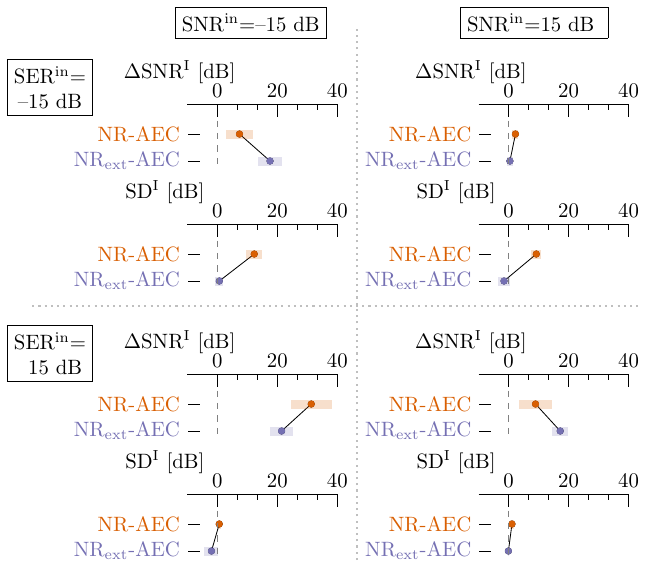}
	\caption{NR performance using the converged filters for the entire data. Dots show mean performance and shading the standard deviation. At low SER\textsuperscript{in}, NR\textsubscript{ext}-AEC has better NR performance as the NR\textsubscript{ext} filters scale with the number of loudspeakers opposed to the NR filters. Only at high SER\textsuperscript{in} and low SNR\textsuperscript{in}, NR-AEC has better NR performance as the NR filters use a lower rank-approximation than the NR\textsubscript{ext} filters, limiting the noise from each mode.}
	\label{fig:comparison}
\end{figure}

\begin{figure}
	\centering
	\includegraphics[width=0.902\linewidth]{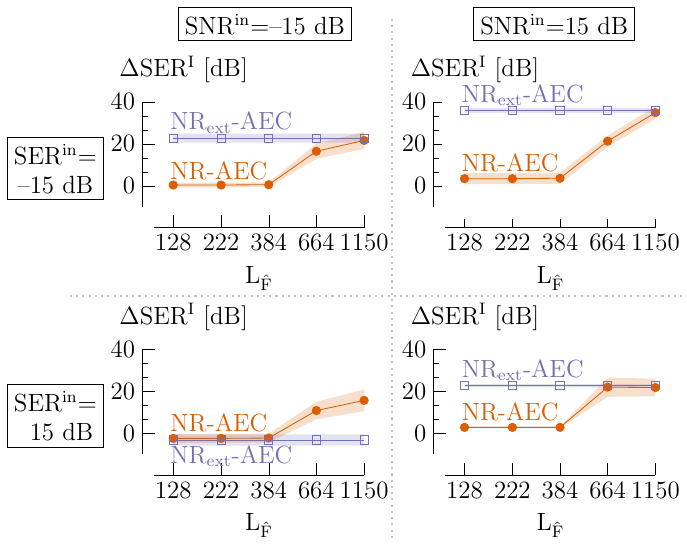}
	\caption{AEC performance in function of $L_{\hat{F}}$ using the converged filters for the entire data. Dots show mean performance and shading the standard deviation. As the AEC filters in NR\textsubscript{ext}-AEC are independent of the NR\textsubscript{ext} filters, as opposed to NR-AEC, NR\textsubscript{ext}-AEC performance exceeds NR-AEC. Only at high SER\textsuperscript{in} and low SNR\textsuperscript{in} the top performance is higher in NR-AEC, yet this advantage is lost with adaptive filters (Fig. \ref{fig:seri_adaptive}).}
	\label{fig:seri}
\end{figure}

Fig. \ref{fig:comparison} and Fig. \ref{fig:seri} show the NR and AEC performance of NR-AEC and NR\textsubscript{ext}-AEC as a function of $L_{\hat{F}}$ when using the converged filters for the entire data. As $L_{\hat{F}}$ does not influence the NR and NR\textsubscript{ext} filters, Fig. \ref{fig:comparison} is independent of $L_{\hat{F}}$.

Regarding the NR performance, when SER\textsuperscript{in} is lower than or comparable to SNR\textsuperscript{in}, NR\textsubscript{ext}-AEC attains a better NR performance than NR-AEC. Indeed, the NR can be interpreted as an MWF aiming at cancelling the one near-end room noise and two echo sources. However, as there are two microphones, the NR filters only have one degree of freedom, which is insufficient to cancel the two dominant echo signals, resulting in a low SNR\textsuperscript{I} improvement and high SD\textsuperscript{I} degradation. NR\textsubscript{ext} does not suffer from this phenomenon, as the degrees of freedom in the NR\textsubscript{ext} filters scale with the number of loudspeakers, thus attaining a larger SNR\textsuperscript{I} improvement and lower SD\textsuperscript{I} degradation. Only at high SER\textsuperscript{in} combined with low SNR\textsuperscript{in}, better NR performance is achieved in NR-AEC. As NR\textsubscript{ext} uses a larger rank-approximation in the GEVD ($3$ rather than $1$), noise along an increased number of modes in the GEVD is retained in the NR\textsubscript{ext} filters than in the NR filters.

\begin{figure}
	\centering
	\includegraphics[width=0.902\linewidth]{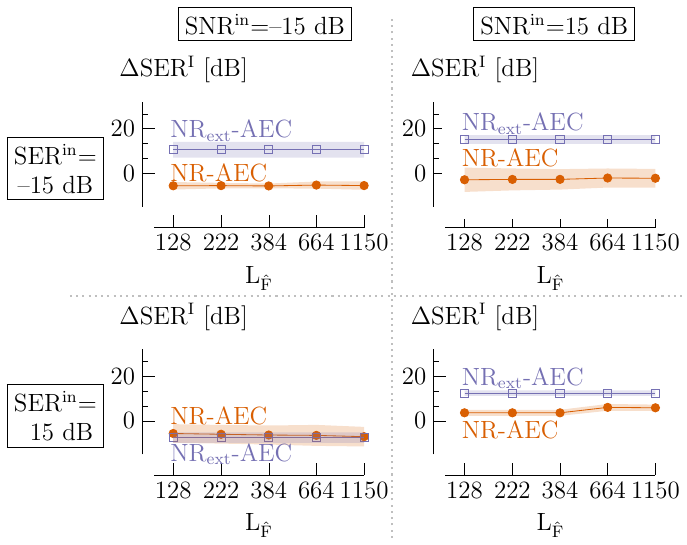}
	\caption{AEC performance in function of $L_{\hat{F}}$, when adapting the filters through time. NR-AEC shows decreased performance compared to Fig. \ref{fig:seri} as the AEC filters in NR-AEC need to track the adaptivity of the NR filters.}
	\label{fig:seri_adaptive}
\end{figure}

NR\textsubscript{ext}-AEC also attains a better AEC performance than NR-AEC at lower $L_{\hat{F}}$ in Fig. \ref{fig:seri}. The AEC filters in NR\textsubscript{ext}-AEC only need to model the $128$ coefficient long echo path impulse responses, while the AEC filters in NR-AEC need to model the $1150$ coefficients in the convolution of the echo path impulse responses and the NR filters. Only at high SER\textsuperscript{in} combined with low SNR\textsuperscript{in} the top AEC performance of NR-AEC exceeds that of NR\textsubscript{ext}-AEC as the AEC filters in NR-AEC are less perturbed by the noise; and as imperfect NR\textsubscript{ext} filters do influence the AEC filters to some extent, but these NR\textsubscript{ext} filters cannot be trivially modelled by the AEC filters as the NR\textsubscript{ext} filters also alter the loudspeaker signals. However, when considering adaptive filters (Fig. \ref{fig:seri_adaptive}), this advantage of NR-AEC is ineffective, as the AEC filters then need to track the adaptivity of the NR filters, unlike in NR\textsubscript{ext}-AEC.

\section{Conclusion}\label{section:conclusion}
We have introduced a cascaded design for combined NR and AEC under the assumption of the echo paths being additive maps, thus preserving the addition operation. To this end, we have proposed an NR\textsubscript{ext} preceding an AEC (NR\textsubscript{ext}-AEC), and compared this NR\textsubscript{ext}-AEC to the traditional design with an NR preceding an AEC (NR-AEC). Whereas NR filters aim at reducing the near-end room noise (and possibly partially the echo) and operate on microphones only, the NR\textsubscript{ext} filters aim at reducing both the near-end room noise and the far-end room noise component in the echo, and operate on both the microphones and the loudspeakers. NR\textsubscript{ext}-AEC outperforms NR-AEC in terms of both the NR and AEC performance, as the AEC filters in NR\textsubscript{ext}-AEC remarkably become independent of the NR\textsubscript{ext} filters whereas the AEC filters in NR-AEC are dependent on the NR filters, and as the NR\textsubscript{ext} filters scale with the number of loudspeakers while the NR filters do not.
\bibliographystyle{IEEEtran}
\bibliography{ref}

\begin{thebibliography}{10}
\providecommand{\url}[1]{#1}
\csname url@samestyle\endcsname
\providecommand{\newblock}{\relax}
\providecommand{\bibinfo}[2]{#2}
\providecommand{\BIBentrySTDinterwordspacing}{\spaceskip=0pt\relax}
\providecommand{\BIBentryALTinterwordstretchfactor}{4}
\providecommand{\BIBentryALTinterwordspacing}{\spaceskip=\fontdimen2\font plus
\BIBentryALTinterwordstretchfactor\fontdimen3\font minus
  \fontdimen4\font\relax}
\providecommand{\BIBforeignlanguage}[2]{{%
\expandafter\ifx\csname l@#1\endcsname\relax
\typeout{** WARNING: IEEEtran.bst: No hyphenation pattern has been}%
\typeout{** loaded for the language `#1'. Using the pattern for}%
\typeout{** the default language instead.}%
\else
\language=\csname l@#1\endcsname
\fi
#2}}
\providecommand{\BIBdecl}{\relax}
\BIBdecl

\bibitem{hanslerTopicsAcousticEcho2006}
E.~H{\"a}nsler and G.~Schmidt, \emph{Topics in {{Acoustic Echo}} and {{Noise
  Control}}}.\hskip 1em plus 0.5em minus 0.4em\relax Berlin, Heidelberg:
  Springer-Verlag, 2006.

\bibitem{serizelLowrankApproximationBased2014}
R.~Serizel, M.~Moonen, B.~Van~Dijk, and J.~Wouters, ``Low-rank {{Approximation
  Based Multichannel Wiener Filter Algorithms}} for {{Noise Reduction}} with
  {{Application}} in {{Cochlear Implants}},'' \emph{IEEE/ACM Transactions on
  Audio, Speech, and Language Processing}, vol.~22, no.~4, pp. 785--799, Apr.
  2014.

\bibitem{gustafssonCombinedAcousticEcho1998a}
S.~Gustafsson, R.~Martin, and P.~Vary, ``Combined acoustic echo control and
  noise reduction for hands-free telephony,'' \emph{Signal Processing},
  vol.~64, no.~1, pp. 21--32, Jan. 1998.

\bibitem{docloCombinedAcousticEcho2000}
S.~Doclo, M.~Moonen, and E.~{de Clippel}, ``Combined acoustic echo and noise
  reduction using {{GSVD-based}} optimal filtering,'' in \emph{2000 {{IEEE
  International Conference}} on {{Acoustics}}, {{Speech}}, and {{Signal
  Processing}} ({{ICASSP}})}, Istanbul, Turkey, Jun. 2000, pp. II1061--II1064
  vol.2.

\bibitem{cohenJointBeamformingEcho2018}
A.~Cohen, A.~Barnov, S.~{Markovich-Golan}, and P.~Kroon, ``Joint
  {{Beamforming}} and {{Echo Cancellation Combining QRD Based Multichannel
  AEC}} and {{MVDR}} for {{Reducing Noise}} and {{Non-Linear Echo}},'' in
  \emph{2018 26th {{European Signal Processing Conference}} ({{EUSIPCO}})},
  Rome, Italy, Sep. 2018, pp. 6--10.

\bibitem{luisvaleroLowComplexityMultiMicrophoneAcoustic2019a}
M.~Luis~Valero and E.~A.~P. Habets, ``Low-{{Complexity Multi-Microphone
  Acoustic Echo Control}} in the {{Short-Time Fourier Transform Domain}},''
  \emph{IEEE/ACM Transactions on Audio, Speech, and Language Processing},
  vol.~27, no.~3, pp. 595--609, Mar. 2019.

\bibitem{martinAcousticEchoCancellation1997}
R.~Martin, S.~Gustafsson, and M.~Moser, ``Acoustic echo cancellation for
  microphone arrays using switched coefficient vectors,'' in \emph{1997
  {{International Workshop}} on {{Acoustic Echo}} and {{Noise Control}}
  ({{IWAENC}}),}, London, United Kingdom, 1997, pp. I--85.

\bibitem{schrammenChangePredictionLow2019}
M.~Schrammen, A.~Bohlender, S.~K{\"u}hl, and P.~Jax, ``Change {{Prediction}}
  for {{Low Complexity Combined Beamforming}} and {{Acoustic Echo
  Cancellation}},'' in \emph{2019 27th {{European Signal Processing
  Conference}} ({{EUSIPCO}})}, A Coruna, Spain, Sep. 2019, pp. 1--5.

\bibitem{reuvenJointAcousticEcho2004}
G.~Reuven, S.~Gannot, and I.~Cohen, ``Joint acoustic echo cancellation and
  transfer function {{GSC}} in the frequency domain,'' in \emph{23rd {{IEEE
  Convention}} of {{Electrical}} and {{Electronics Engineers}} in {{Israel}}},
  Tel-Aviv, Israel, Sep. 2004, pp. 412--415.

\bibitem{roebbenGithubRepositoryCascaded2024}
A.~Roebben, ``Github repository: {{Cascaded}} noise reduction and acoustic echo
  cancellation based on an extended noise reduction,''
  https://github.com/Arnout-Roebben/NRAEC\_vs\_NRextAEC, 2024.

\bibitem{ruizCascadeAlgorithmsCombined2023a}
S.~Ruiz, T.~{van Waterschoot}, and M.~Moonen, ``Cascade {{Algorithms}} for
  {{Combined Acoustic Feedback Cancellation}} and {{Noise Reduction}},''
  \emph{EURASIP Journal on Audio, Speech, and Music Processing}, vol. 2023,
  no.~1, p.~37, Sep. 2023.

\bibitem{sprietAdaptiveFilteringTechniques2004}
A.~Spriet, ``{Adaptive filtering techniques for noise reduction and acoustic
  feedback cancellation in hearing aids},'' Ph.D. dissertation, KU Leuven,
  Leuven, Belgium, 2004.

\bibitem{romboutsAdaptiveFilteringAlgorithms2003}
G.~Rombouts, ``Adaptive filtering algorithms for acoustic echo and noise
  cancellation,'' Ph.D. dissertation, KU Leuven, Leuven, Belgium, 2003.

\bibitem{docloMultimicrophoneNoiseReduction2003}
S.~Doclo, ``{Multi-microphone noise reduction and dereverberation techniques
  for speech applications},'' Ph.D. dissertation, KU Leuven, Leuven, Belgium,
  2003.

\bibitem{luInversesBlockMatrices2002}
T.-T. Lu and S.-H. Shiou, ``Inverses of 2 {\texttimes} 2 block matrices,''
  \emph{Computers \& Mathematics with Applications}, vol.~43, no. 1-2, pp.
  119--129, Jan. 2002.

\bibitem{desenaModelingRectangularGeometries2015}
E.~De~Sena, N.~Antonello, M.~Moonen, and T.~{van Waterschoot}, ``On the
  {{Modeling}} of {{Rectangular Geometries}} in {{Room Acoustic
  Simulations}},'' \emph{IEEE/ACM Transactions on Audio, Speech, and Language
  Processing}, vol.~23, no.~4, pp. 774--786, Apr. 2015.

\bibitem{nilssonDevelopmentHearingNoise1994}
M.~Nilsson, S.~D. Soli, and J.~A. Sullivan, ``Development of the {{Hearing In
  Noise Test}} for the measurement of speech reception thresholds in quiet and
  in noise,'' \emph{The Journal of the Acoustical Society of America}, vol.~95,
  no.~2, pp. 1085--1099, Feb. 1994.

\bibitem{auditecAuditoryTestsRevised1997}
Auditec, ``Auditory {{Tests}} ({{Revised}}), {{Compact Disc}}, {{Auditec}},''
  St. Louis, MO, 1997.

\bibitem{didierSamplingRateOffset2023}
P.~Didier, T.~{van Waterschoot}, S.~Doclo, and M.~Moonen, ``Sampling {{Rate
  Offset Estimation}} and {{Compensation}} for {{Distributed Adaptive
  Node-Specific Signal Estimation}} in {{Wireless Acoustic Sensor Networks}},''
  \emph{IEEE Open Journal of Signal Processing}, vol.~4, pp. 71--79, 2023.

\bibitem{avargelSystemIdentificationShortTime2007a}
Y.~Avargel and I.~Cohen, ``System {{Identification}} in the {{Short-Time
  Fourier Transform Domain With Crossband Filtering}},'' \emph{IEEE
  Transactions on Audio, Speech, and Language Processing}, vol.~15, no.~4, pp.
  1305--1319, May 2007.

\bibitem{zhuRobustLightweightVoice2023b}
Z.~Zhu, L.~Zhang, K.~Pei, and S.~Chen, ``A robust and lightweight voice
  activity detection algorithm for speech enhancement at low signal-to-noise
  ratio,'' \emph{Digital Signal Processing}, vol. 141, p. 104151, Sep. 2023.

\bibitem{greenbergIntelligibilityWeightedMeasures1993}
J.~E. Greenberg, P.~M. Peterson, and P.~M. Zurek, ``Intelligibility-weighted
  measures of speech-to-interference ratio and speech system performance,''
  \emph{The Journal of the Acoustical Society of America}, vol.~94, no.~5, pp.
  3009--3010, Nov. 1993.

\end{thebibliography}

\end{document}